# Electrochemical removal of antibiotics and multi-drug resistant bacteria using S-functionalized graphene sponge electrodes

*Natalia Ormeño Cano*[a,b], *Jelena Radjenovic* [a,c*]

[a]*Catalan Institute for Water Research (ICRA-CERCA), c/Emili Grahit, 101, 17003 Girona, Spain*

[b]*University of Girona, Girona, Spain*

[c]*Catalan Institution for Research and Advanced Studies (ICREA), Passeig Lluís Companys 23, 08010 Barcelona, Spain*

** Corresponding author:*

*Jelena Radjenovic, Catalan Institute for Water Research (ICRA), Scientific and Technological Park of the University of Girona, 17003 Girona, Spain*

Phone: + 34 972 18 33 80; Fax: +34 972 18 32 48; E-mail: jradjenovic@icra.cat

## Abstract

In this study, we synthesized S-functionalized graphene sponge electrode and applied it for electrochemical oxidation of five commonly used antibiotics, namely sulfamethoxazole, trimethoprim, ofloxacin, roxithromycin and erythromycin, and the inactivation of a multi-drug resistant *Escherichia coli* (*E. coli*). The experiments were performed using real drinking water in a flow-through, one-pass mode. Highly polar antibiotics such as sulfamethoxazole did not adsorb onto the graphene sponge but were completely removed (i.e., ≥95% removal) at low applied current densities (14.5 A $m^{-2}$). Antibiotics with high affinity for π-π interactions such as ofloxacin were completely removed already in the open circuit, and current application led to their further degradation. S-doped graphene sponge anode resulted in 4.5 log removal of a multi-drug resistant *E. coli* at 29 A $m^{-2}$. There was no regrowth of bacteria observed during storage of the electrochemically treated samples, suggesting that the treatment severely impacted the cell viability. Further *E. coli* removal of 0.7 log was observed after the storage of electrochemically treated samples. The energy consumption of a continuously operated electrochemical system that achieved 4.5 log inactivation of a multi-drug resistant *E. coli* and 87 - 99% removal of antibiotics was 1.1 kWh $m^{-3}$.

## Introduction

The indiscriminate use of antibiotics in human medicine, animal husbandry and aquaculture has promoted an unprecedented spread of antimicrobial resistance worldwide (Zhang et al., 2022). Antimicrobials are only partially metabolized and the rest is excreted, thus they are frequently detected in different water sources such as groundwater, soil, surface water and even in treated drinking water (Jia et al., 2017; Wang et al., 2021, 2016). Continuous exposure and ubiquitousness of antibiotics contributes to the proliferation and prevalence of antibiotic-resistant bacteria (ARB) and antibiotic-resistance genes (ARG) in the environment (Pruden et al., 2013). The spread of ARBs and ARGs is one of the mayor global threats of the 21st century that currently causes at least 700,000 deaths worldwide each year (World Health Organization, 2019). Therefore, to ensure a "universal and equitable access to safe and affordable drinking water for all", innovative low-cost water treatment and disinfection technologies are needed, capable of inactivating antibiotic resistant pathogens, and removing antibiotics and other pollutants associated with the antimicrobial resistance (e.g., antidepressants) (Wang et al., 2023; World Health Organization, 2017).

Electrochemical systems are very well-suited for decentralized and distributed (waste)water treatment due to their modular and easily scalable design, no usage of chemicals, and operation at ambient temperature and pressure (Radjenovic and Sedlak, 2015). However, commercial electrodes used in electrochemical water treatment systems such as boron-doped diamond (BDD), Magnéli phase and mixed metal oxide (MMO) anodes suffer from two major limitations: production of toxic chlorinated byproducts (e.g., organochlorines, chlorate and perchlorate) in presence of chloride, and high cost (Wenderich et al., 2021). On the contrary, recently developed graphene sponge electrodes are characterized by the low electrocatalytic activity towards $Cl^-$ electrooxidation and thus no formation of chlorinated byproducts, and very low estimated production cost (<50 € per $m^2$) (Baptista-Pires et al., 2021). These characteristics make graphene-enabled electrochemical system advantageous compared to conventional electrochemical system, in particular as low-cost point-of-use water treatment unit, given the absence of chlorate and perchlorate formation, and its disinfection performance (Norra et al., 2022; Segues Codina et al., 2023).

One major limitation when electrochemically treating low-conductivity solutions such as drinking water is an increased thickness of the electric double layer, which limits the interaction of trace organic contaminants with the electrode surface (Ormeno-Cano and Radjenovic, 2022). S-functionalization of the reduced graphene oxide (RGO) was previously applied to enhance its surface wettability and increase the number of adsorption sites in supercapacitors and sensors, thus enhancing their electrocatalytic activity (Ma et al., 2022). In this study, graphene sponge electrode was modified with sulfur-containing functional groups and applied for the removal of antibiotics and ARB from real drinking water. S-functionalized graphene sponge (SRGO) anode was coupled with a nitrogen-doped graphene sponge cathode (NRGO) and employed in one pass, flow-through mode for electrochemical degradation. An effective measure to control the spread of ARGs would involve the elimination of both antibiotics and ARBs. For that reason, we evaluated the removal of a set of antibiotics, namely sulfamethoxazole (SMX), ofloxacin (OFX), erythromycin (ERT), trimethoprim (TMP) and roxithromycin (ROX), and for the inactivation of a gram- negative multi-drug resistant *E. coli* carrying genes that confer resistance to previously mentioned SMX and TMP. The main objectives of this study were: *i)* to determine the impact of S-functionalization on the graphene sponge electrode performance, *ii)* to evaluate the removal of antibiotics and ARB in challenging conditions of low-conductivity drinking water, *iii)* to investigate the participation of different oxidant species contributing to electrochemical degradation of pollutants and bacteria inactivation, and *iv)* to investigate the impact of current on their removal efficiencies.

## 2. Materials and methods

### 2.1 Graphene sponge synthesis and characterization

Graphene sponges were synthesized using a previously developed bottom-up method (Baptista-Pires et al., 2021). N-RGO was produced using urea (300 g $L^{-1}$) as an N-source. S-RGO was synthesized based on a modified methodology by Chen et al (Chen et al., 2014), using $Na_2S \cdot 9H_2O$ as a precursor in a ratio of 0.6:1, S:GO (wt:wt) respectively. 5 g $L^{-1}$ of $Na_2S$ nonahydrate (Sigma-Aldrich) was diluted in a commercial GO solution (4 g $L^{-1}$, Graphenea, Spain). This solution was adjusted to neutral pH to avoid $H_2S$ volatilization at acidic pH and prevent $S^{2-}$ formation at alkaline pH conditions. The solution of GO with urea and $Na_2S$ as N and S source, respectively, was used to soak mineral wool template (Diaterm, Spain) and subject to a hydrothermal reaction for 12h at 180°C. The resulting S and N-doped graphene sponges were connected to stain steel current feeders and employed in flow-through reactor as anode and cathode, respectively. To determine the impact of S-functionalization on the graphene sponge anode performance, additional experiments were performed using an undoped graphene sponge anode (RGO). Details of the characterization of the graphene sponge electrodes are given in **Text S1.**

### 2.2 Electrochemical experiments

Multi-channel potentiostat/galvanostat VMP-300 (BioLogic, U.S.A.) was used for running the electrochemical experiments in galvanostatic mode, and for conducting the electrochemical impedance spectroscopy (EIS) experiments. All experiments were conducted in a cylindrical flow–through reactor in one pass mode, at 5 mL $min^{-1}$, corresponding to hydraulic residence time (HRT) of 3.45 min and 175 L $m^{-2}$ $h^{-1}$ (LMH) of the normalized volumetric flux, where the projected surface area of the electrodes was 17.54 $cm^2$. The flow direction was SRGO anode (A) - NRGO cathode (C), given that it was previously determined to be more efficient for the removal of organic contaminants (**Figure S1**) (Baptista-Pires et al., 2021). All experiments were conducted in real drinking water (**Table S1)** having a low electrical conductivity of 0.45 mS $cm^{-1}$, to investigate the system performance under realistic conditions of water treatment, where ohmic drop represents a major challenge. After volatilizing the residual chlorine from the sampled drinking water to exclude its interference in electrochemical disinfection, drinking water was amended with antibiotics at the initial concentration of 0.2 µM, and the *E. coli* strain DSM 103246 carrying antibiotic resistant genes (aph(3=')-lb, aph(six)-ld, aph(3=)-lc, aadA2, blaCTX-M-14b, sul2, sul1, tet(B), and dfrA16) at the initial concentration of ~$10^7$ CFU $mL^{-1}$. The continuously operated flow-through electrochemical reactor was first run in the initial open circuit ($OC_i$), i.e., without applying the current, to evaluate the removal of antibiotics due to adsorption, and bacterial inactivation at the uncharged graphene sponges. Next, electrochemical degradation/inactivation was evaluated in the chronopotentiometric mode at 14.5 and 29 A $m^{-2}$ of anodic current density. At the end of each run, the current was switched off in the final OC ($OC_f$) to confirm that the electrochemical removal of antibiotics and bacteria was not due to their electrosorption only (and therefore accumulation in the system), but also due to their subsequent electro-degradation. For example, an increase in the contaminant concentration in the $OC_f$ above the effluent concentrations observed in the $OC_i$ would indicate that a fraction of a target contaminant was only electrosorbed. To evaluate the impact of the S-functionalization, experiments were also conducted using the undoped graphene sponge anode (RGO) coupled with the NRGO cathode, in an RGO/NRGO system. To confirm the electrochemical inactivation of the bacteria and verify if the induced damage excludes their posterior reactivation, electrochemically treated samples were stored at 37°C, as the optimum temperature for the growth of *E. coli* (Gorito et al., 2021; Noor et al., 2013). All electrode potentials are expressed versus Standard Hydrogen Electrode (/SHE, V). Determination of ohmic drop-corrected potentials, electric energy per order ($E_{eo}$, kWh $m^{-3}$), energy consumption and electrochemically active surface area (EASA) are presented in **Text S2 and S3**. Electrogeneration of hydrogen peroxide ($H_2O_2$), ozone ($O_3$) and hydroxyl radicals (•OH) was evaluated at 14.5 and 29 A $m^{-2}$ of anodic current density. Whereas the formation of •OH and cathodically generated $H_2O_2$ was evaluated in the anode-cathode flow direction employed in the experiments, the formation of anodically generated $O_3$ was evaluated in the cathode-anode flow direction **(Text S4).** All results are presented as means of at least two separate runs with their standard deviations. Details of the analytical methods employed for the analysis of target antibiotics are summarized in **Text S5**.

### 2.3 Microbiological analysis

Multi-drug resistant *E. coli* (DSM 103246) was prepared by growing the bacteria overnight at 37 °C in Luria-Bertani (LB) broth (2.5 g in 100 mL), spiked with antibiotics (6.25 mg $L^{-1}$ of ampicillin) to create a medium selective to antibiotic-resistant strains. The overnight culture with an approximate concentration of $10^8$–$10^9$ CFU $mL^{-1}$ was harvested by centrifuging the broth for 10 min at 3,500 rpm (Eppendorf 5804R). Then, the LB broth supernatant was discarded, and the precipitated bacteria was diluted into chlorine-free drinking water. To determine the multi-drug resistant *E. coli*, an aliquot of the corresponding suspension was diluted 1:10 in a sterile ringer solution up to six-fold serial dilutions. Afterwards, the dilutions were filtered using 0.45 μm pore size cellulose filters (Merck) and plated in Chromocult agar plates. Chromocult agar plates were incubated overnight in dark at 37 °C, and the dark-blue colonies forming units were counted. To determine the removal of *E. coli*, the reactor was operated first in the initial open circuit ($OC_i$), at 14.5 and 29 A $m^{-2}$ of anodic current density, and in the final OC ($OC_f$). To investigate a possible reactivation of *E. coli* cells after the electrochemical treatment, treated samples were incubated overnight in dark at 37°C for 16 h. Samples were diluted in a sterile Ringer solution, filtered, and plated as explained previously, to evaluate a possible regrowth of the inactivated bacteria.

## 3. Results and discussion

### 3.1 Surface and electrochemical characterization of graphene sponge electrodes

The characterization of NRGO sponge employed as cathode is reported in detail in our previous studies (Ormeno-Cano and Radjenovic, 2022). Scanning electron microscopy (SEM) analysis of the SRGO sponge confirmed a complete and homogeneous coating of the template (**Figure S2**). X-ray diffraction (XRD) analysis showed a broad peak 2ϴ at 25°. The shift from the characteristic peak of GO from 11° to 25° is caused by the reduction of GO and doping of RGO, and thus elimination of the oxygen functional groups from the basal plane (**Figure S3**) (Chen et al., 2014). A small peak approximately in 2ϴ at 20° suggest that a fraction of GO is not completely intercalated, and the GO is not fully interconnected with the oxygen atoms.(Hsiao et al., 2010; Mundinamani, 2020) X-ray photoelectron spectroscopy (XPS) analysis confirmed that the total content of atomic S was 1.3 % for SRGO (**Figure 1C, B, Table S3**). The S2p spectrum of SRGO (**Figure 1C**) can be resolved into four different peaks with binding energies of 163.3 eV (7.3%, -C-S-C-), and 168.4 eV (41.9%), 170.5 eV (42.54%) and 172.8 eV (8.4%), the last three attributed to -C-SOx-C- groups (Hussain et al., 2015; Pan et al., 2022; C. Wang et al., 2015; Wang et al., 2020).

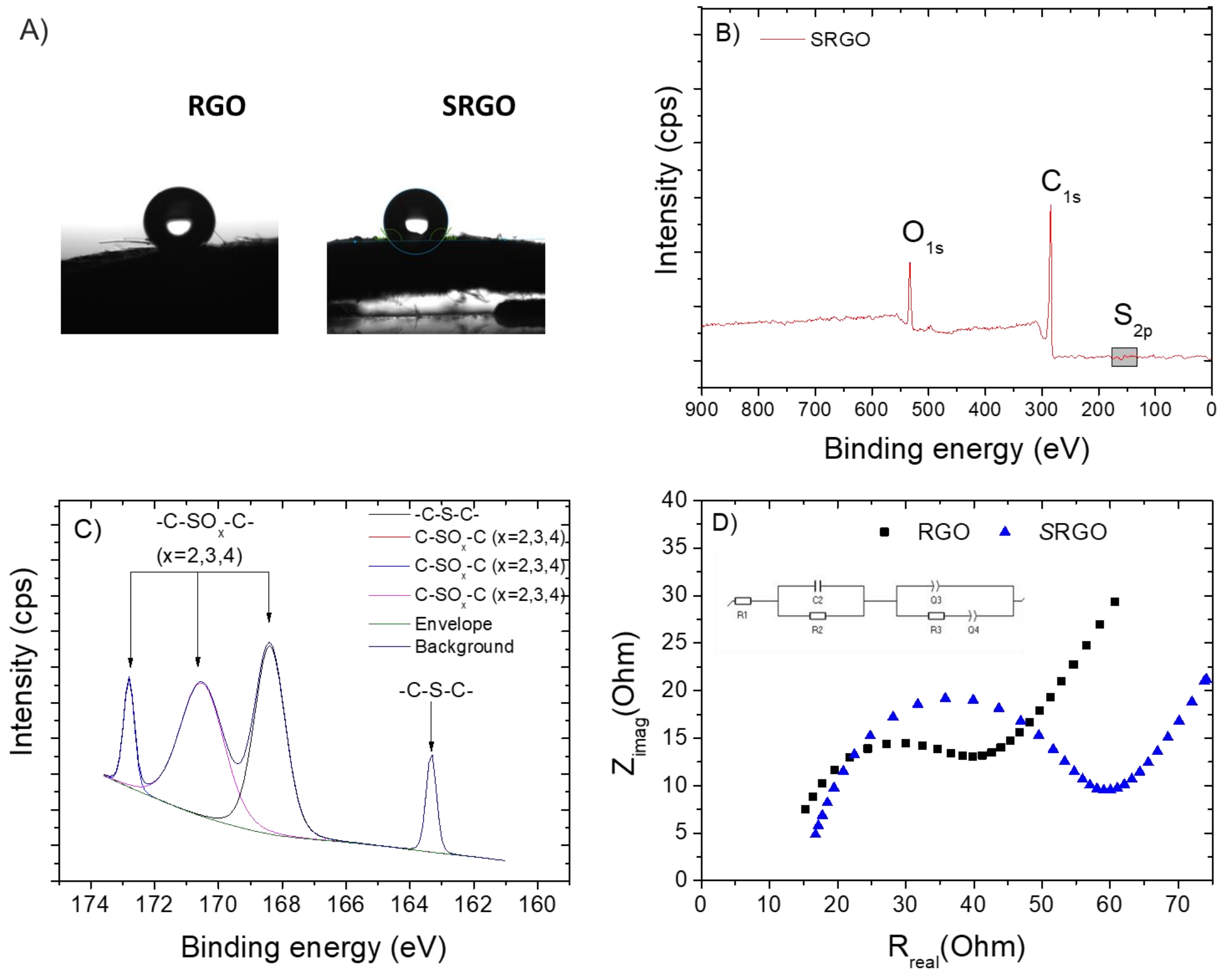

**Figure 1. A)** Contact angle of RGO and SRGO sponges **B)** XPS spectra of SRGO C**)** S2p spectra of SRGO **D)** Nyquist plots of RGO and SRGO in the frequency range of 5MHz - 1Hz.

Furthermore, XPS analysis indicated the C/O atomic ratio was 2.23 for SRGO. Given that the C/O ratio of the undoped RGO sponge was determined to be 2.51 (Baptista-Pires et al., 2021), this result suggest lower reduction degree and/or enhanced presence of oxygen functionalities in the graphene sponge electrode functionalized with sulfur. Given the previously explained spectra of atomic S, most of the S-containing functional groups were -C-$SO_X$-C- (92.7%) (**Table S4**), which may explain the somewhat lower C/O ratio of SRGO in comparison with RGO. In the C1s spectrum, C=C (284.4 eV, 33%) and π-π (290.5 eV, 4.4%) bonds were identified, as well as C-O/C-S and C=O peaks located at 286.1 and 287.8 eV (**Table S4 and Figure S4A**) (Poh et al., 2013; C. Wang et al., 2015). The appearance of C-S bond at 286.1 eV confirmed the covalent bonding of the atomic S into the graphene lattice (Poh et al., 2013). For C1s in SRGO (**Table S4**), π -π interactions in carbon materials correspond to an aromatic or unsaturated structure (Q. Zhang et al., 2020). The O1s spectrum of SRGO was fitted into three peaks at 531.3, 533.3 and 534.3 eV, assigned to C-O, C=O and O=C-O bonds (**Figure S4B and Table S4)** (Baptista-Pires et al., 2021; Zhang et al., 2018).

The ζ potential of the SRGO sponge was determined to be -29 mV, more negative compared with the previously determined ζ potential of RGO (-18.1 mV) (Cuervo Lumbaque et al., 2022). This can be explained by the presence of more negatively charged oxygen functional groups (e.g., -C-O, C=O- and O=C-O-, 30.6 %, **Table S3 and S4**) and sulphur oxide groups (e.g., -C-$SO_X$-) (Li et al., 2015; Ollik et al., 2021).

To compare the impact of the S-functionalization on the electrical resistance of graphene sponge electrode, EIS analysis was conducted for SRGO and RGO electrodes (**Figure 1D**, **Table S5**). Ohmic resistances ($R_\Omega$) of RGO and SRGO anodes were very similar, 13.5 and 13.7 **Ω**, respectively, as this resistance represents the potential drop between the reference electrode and the anode and depends on the ionic strength of the supporting electrolyte. The charge transfer resistance ($R_{ct}$) and double layer resistance at the electrode-electrolyte interface is represented by the diameter of the semi-circle in the region of high- to mid-frequency (Chen et al., 2014; Cuervo Lumbaque et al., 2022). The straight line in this frequency region represents the electrochemical process governed by the mass transfer phenomena at the electrode surface and can differ depending on $R_{CT}$ and double layer capacitance ($C_{dl}$) parameters (Lazanas and Prodromidis, 2023; Lumbaque and Radjenovic, 2023). As can be observed from **Figure 1D**, S-functionalization of RGO increased the $R_{CT}$ from 48.7 to 79 Ω. This is due to the poor electrical conductivity of elemental sulfur which is partially introduced in the porous carbon matrix (Kiciński et al., 2014). Moreover, sulfur is also a larger atom (100 pm) and may disrupt the carbon connection pattern, creating more defects in the graphene matrix and resulting in a disordered graphene lattice, which decreases the material conductivity (Wohlgemuth et al., 2012b). Nevertheless, the same defects introduced by the addition of S-dopant can also promote the electrocatalytic activity of the material, even though its electrical conductivity is lowered (Wang and Han, 2022; Wohlgemuth et al., 2012a; X. Zhang et al., 2020).

EASA determined for the SRGO electrode was 0.97 $m^2$ $g^{-1}$, somewhat higher compared with the RGO electrode (0.81 $m^2$ $g^{-1}$) (Cuervo Lumbaque et al., 2022). Larger EASA of SRGO enhances the access of the electrolyte ions to the electrode material surface (Ali et al., 2015; Lazanas and Prodromidis, 2023). The calculated value of $C_{dl}$ of SRGO was 0.097 F $g^{-1}$, higher compared with the RGO (0.067 F $g^{-1}$) (Cuervo Lumbaque et al., 2022). Higher $C_{dl}$ of SRGO indicates an improved contact of the SRGO anode and electrolyte, as electrode surface area accessibility is dependent on the double layer capacitance of the material (Deheryan et al., 2014). This is of crucial importance when working with the low-conductivity solutions such as the employed drinking water**.** The contact angles determined for RGO and SRGO were 138.64±4.5º, and 129.9±4.7º, respectively. Therefore, the incorporation of sulfur functionalities led to a significantly lower contact angle, also compared to a boron-doped graphene sponge (BRGO) anode employed for the degradation of antibiotics in our previous study (139.67 ± 4.5°) (Baptista-Pires et al., 2021; Ormeno-Cano and Radjenovic, 2022). Lower contact angle and thus enhanced wettability can be explained by the presence of additional hydrophilic functional groups such as -C-$SO_x$-C- in the RGO coating (Deheryan et al., 2014).

### 3.2 Electrochemical removal of multi-drug resistant *E. coli*

In the absence of current, both SRGO and RGO anodes coupled with the NRGO cathode resulted in the same 1.9 log removal of *E. coli* during the $OC_i$ (**Figure 2, Table S6**). Inactivation of the bacteria at the ggraphene-based coatings occurs *via* membrane stress induced by their contact with the sharp graphene nanosheets that damage the cell membrane (Perreault et al., 2015).

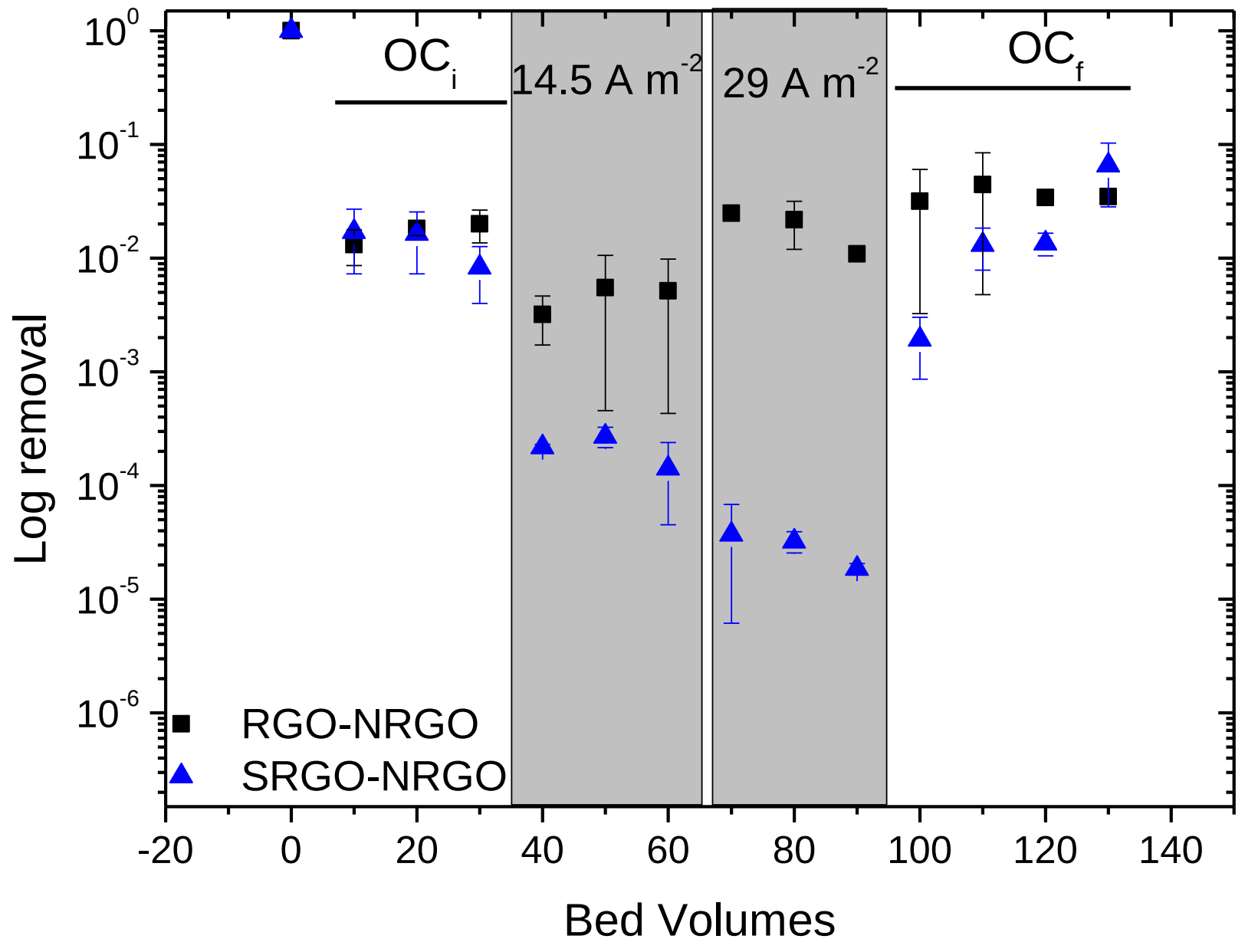

**Figure 2**. Removal of *E. coli* from drinking water using RGO and SRGO anodes coupled to NRGO cathode, during initial open circuit ($OC_i$), final open circuit ($OC_f$), and at 14.5 and 29 A $m^{-2}$ of applied current density.

Electrochemical disinfection was evaluated using a continuous current application of 14.5 A $m^{-2}$ and 29 A $m^{-2}$, resulting in ohmic drop-corrected anode potentials of 1.6 and 1.7 V/SHE for RGO anode, and 2.9 and 3.1 V/SHE for SRGO anode, respectively (**Table S5**). The SRGO anode demonstrated stable performance as there was no increase in the sulfate concentrations in the treated effluent. Application of 14.5 A $m^{-2}$ led to 2.3 log removal of *E. coli* when using RGO anode, and 3.7 log removal with the SRGO anode. In our previous studies, electroporation and irreversible damage of the cell membrane was identified as a determining inactivation mechanism in chlorine-free electrochemical systems equipped with graphene sponge electrodes (Norra et al., 2022; Segues Codina et al., 2023). Although electroporation typically requires large potential gradients (i.e., $>10^7$ V $m^{-1}$) and thus high energy consumption ($>2.710^{-3}$ kW$L^{-1}$) (Huo et al., 2018, 2016; Zhou et al., 2020), it occurs in the case of graphene sponge electrodes at much lower potential gradients (i.e., 2.6-2.8 mV $cm^{-1}$) due to their nanostructured coating that enhances the electric field strength at the electrode surface. Similar phenomena of *E. coli* inactivation via low voltage electroporation was previously reported for carbon nanotube (CNT) sponge electrodes (Huo et al., 2017, 2018). At 14.5 A $m^{-2}$, the potential gradients of RGO/NRGO and SRGO/NRGO systems are very similar, 2.8 and 2.6 V $cm^{-1}$, respectively. Thus, higher log removal of *E. coli* in the SRGO/NRGO system is a consequence of enhanced bacterial interaction with the SRGO anode surface, thus promoting the inactivation of *E. coli*. Further increase in the anodic current density to 29 A $m^{-2}$ resulted in somewhat lower *E. coli* removal for the RGO anode, 1.7 log, likely due to the more intense oxygen evolution reaction and thus obstruction of the anode surface by the formed oxygen bubbles. In the case of SRGO anode, higher current density increased the *E. coli* removal to 4.5 log (**Figure 2, Table S6**). In addition to electroporation, electrogenerated oxidants such as $\cdot OH$, $O_3$ and $H_2O_2$ contribute to electrochemical inactivation of *E. coli* (Cho et al., 2004; Ferro et al., 2017). Previous studies reported the ability of oxidant species to diffuse towards the outer layers of the bacterial cells breaking the cell walls and consequently causing the leaking of the intracellular material and cell death (Bruguera-Casamada et al., 2016; Hunt and Mariñas, 1999). The steady-state concentrations of •OH and cathodically generated $H_2O_2$ were similar for the RGO/NRGO and SRGO/NRGO systems at 29 A $m^{-2}$ s, $4.7 \pm 0.2 * 10^{-14}$ and $4.1 \pm 0.05 * 10^{-14}$ M, and 0.65±0.03 and 0.58±0.04 mg $L^{-1}$, respectively (**Table S7**). However, anodically generated $O_3$ was measured in higher concentrations in the SRGO/NRGO system (0.5±0.01 mg $L^{-1}$) compared with the RGO/NRGO (0.16±0.01 mg $L^{-1}$). It is important to note that $H_2O_2$ is activated to •OH radicals at the N-active sites of NRGO cathode (Su et al., 2019); thus, the measured amount is a residual amount of $H_2O_2$ in the system. Higher concentration of oxidant species contributed to an enhanced removal of *E. coli* in the SRGO/NRGO system (i.e., 4.5 log) compared with the RGO/NRGO system (1.7 log at 29 A $m^{-2}$).

To verify that the inactivation of bacteria was irreversible, electrochemically treated samples were stored overnight at 37 °C and evaluated for bacterial regrowth (**Figure 3**). Storage of the samples from the RGO/NRGO system led to a further inactivation of *E. coli* and an additional 0.2 log removal at both 14.5 and 29 A $m^{-2}$, whereas in the case of the SRGO/NRGO system, sample storage resulted in an additional *E. coli* removal of 0.4 (14.5 A $m^{-2}$) and 0.7 log (29 A $m^{-2}$) (**Figure 3**). Higher inactivation of *E. coli* during storage in the case of SRGO/NRGO system is a consequence of more pronounced cell damage via electroporation during the application of current, as explained in the previous paragraph. In the $OC_f$, gradual return of the *E. coli* concentrations to the initial values excludes the possibility of their accumulation in the system and confirms that the *E. coli* removed during the application of current were not only electrosorbed, but also inactivated.

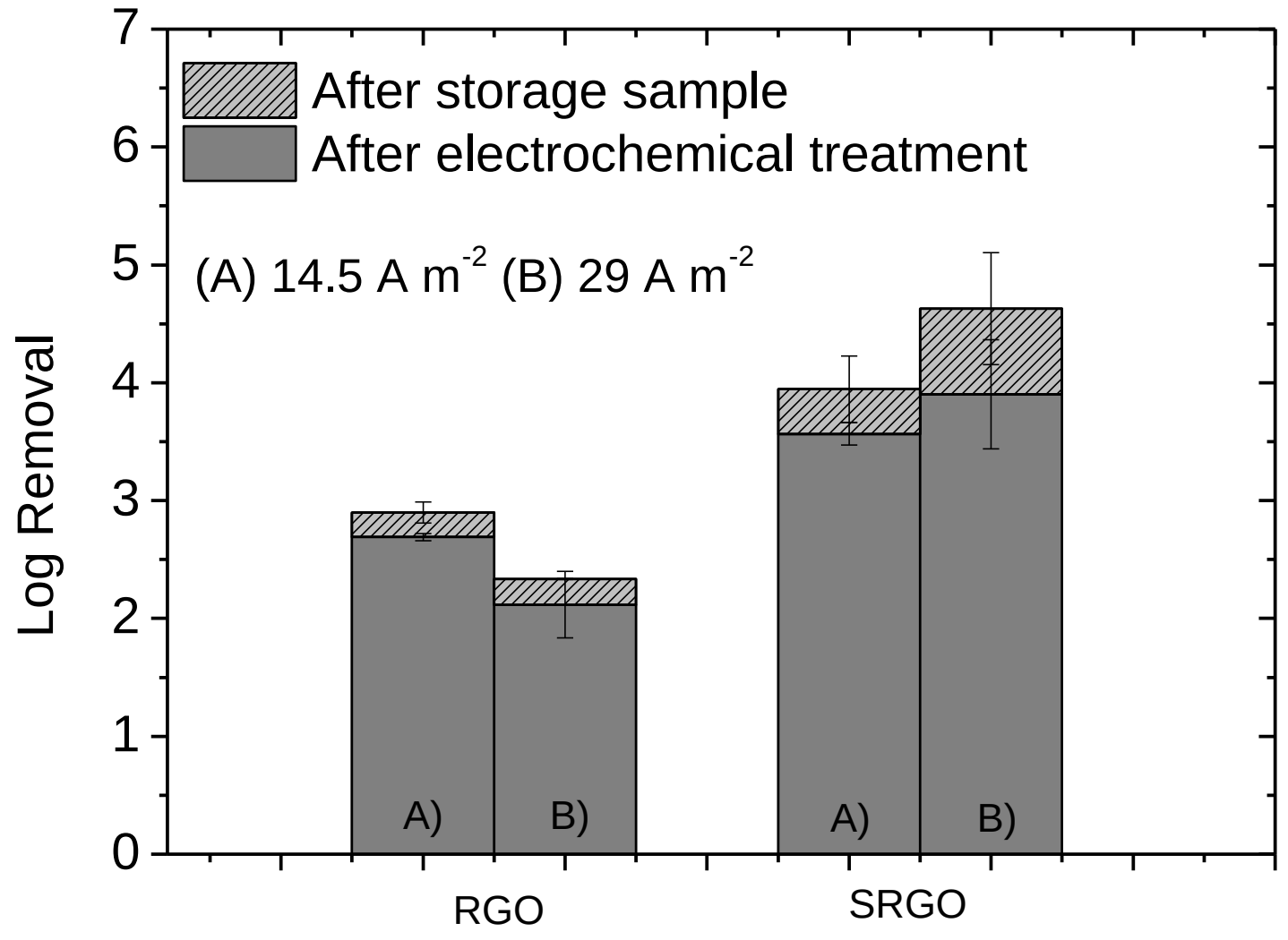

**Figure 3.** Removal of *E. coli* using RGO and SRGO anode versus NRGO cathode determined immediately after sampling the effluent (i.e., “after electrochemical oxidation treatment”), and after storing the effluent sample at 37 °C for 16 h (i.e., “after storage”), at: **A)** 14.5 **B)** and 29 A $m^{-2}$ of applied current density in drinking water.

Thus, the overall removal of multi-drug resistant *E. coli* from drinking water was 1.9 and 5.2 log for the RGO/NRGO and SRGO/NRGO systems operating at 29 A $m^{-2}$, and with the corresponding energy consumption of 1.25 and 1.1 kWh $m^{-3}$, respectively. This result compares favourably with the state-of-the-art on the electrochemical elimination of pathogens in both presence and absence of chlorine (**Table S8**). For example, electrochemical disinfection system with a Magnéli phase $Ti_4O_7$ anode achieved 6.2 log removal of multi-drug resistant *Salmonella enterica*, with an energy consumption of 2.3 kWh $m^{-3}$ but in a higher conductivity medium (**Table S8**) (Wang et al., 2021). Inactivation of 5 log of *E. coli* required 0.27 kWh $m^{-3}$ using $Mo_2C$ loaded on SS as an anode, however in a highly conductive supporting electrolyte (PBS) and in the presence of chloride, which led to the electro-generation of chlorine and thus enhanced disinfection (Liu et al., 2022). Finally, further studies are needed to evaluate the capacity of our system for the elimination of antibiotics resistance genes (ARGs).

### 3.3 Electrochemical removal of antibiotics from drinking water

**Figure 4** presents the observed removals of the target antibiotics in the SRGO/NRGO system at two applied current densities, i.e., 14.5 and 29 A $m^{-2}$. To compare the impact of S-functionalization on the electrochemical degradation of antibiotics, experiments were also performed with the RGO/NRGO system (**Figure S5**). SRGO anode significantly outperformed the undoped graphene sponge anode, resulting in higher removal efficiencies in both $OC_i$ and with the application of current (**Table S9**). The only compound not adsorbed onto the graphene sponge electrodes in the $OC_i$ was SMX, which can be explained by its high hydrophilicity (log D= −0.56, **Table S10**) and negative charge at neutral pH ($pKa_1$=1.4, $pKa_2$=5.8, **Table S10**), thus leading to electrostatic repulsion with the graphene sponge surface. In the case of other antibiotics, the presence of S-functional groups significantly enhanced their removal in the $OC_i$. For example, macrolide antibiotics ERT and ROX adsorbed only partially onto the undoped graphene sponge anode, resulting in less than 36% removal in the $OC_i$. In the case of SRGO anode, ERT was nearly completely adsorbed in the $OC_i$ (92% removal), whereas the removal of ROX was 73%, indicating a pronounced enhancement of the interaction of the target antibiotics with the graphene sponge electrode functionalized with -C-$SO_x$-C- and -C-S-C- groups (**Table S4**). TMP exhibited similar behaviour, with 42% removal in the $OC_i$ in the RGO/NRGO system (**Figure S5**), and 63% removal in the SRGO/NRGO system (**Figure 4**). OFX was completely adsorbed onto the graphene sponge electrodes regardless of the presence of S-functional groups, with >98% removal efficiency in $OC_i$, (**Table S9**). The adsorption of OFX onto the graphene-based materials is governed by the strong π-π electron donor-acceptor interactions, due to the presence of multiple double bonds and aromatic rings in the OFX molecule (**Table S9**) (Ehtesabi et al., 2019), as well as hydrophobic interaction, hydrogen bonding and electron-donor–acceptor and cation- π interactions (Ehtesabi et al., 2019). Cation-π interactions are also the dominant interaction mechanisms for the positively charged macrolide antibiotics (i.e., ROX and ERT), as they do not contain aromatic rings and thus do not interact via π-π stacking (Ehtesabi et al., 2019). Highly polar TMP (log D -1.15) was partially adsorbed onto the graphene sponge electrodes, with 36% and 63% removal observed in the $OC_i$ in the RGO/NRGO and SRGO/NRGO systems, respectively. TMP is present as an uncharged molecule at neutral pH, and can undergo π-π stacking with the RGO coating (Carrales-Alvarado et al., 2020). The pronounced enhancement in the adsorption of the target antibiotics onto the S-functionalized graphene sponge electrodes can be explained by the presence of -C-$SO_x$-C- groups in graphene, which increase the graphene sponge surface hydrophilicity due to the formation of hydrogen bounds between the oxygen atom in $SO_x$ and the hydrogen atom in $H_2O$ molecules (Tavakol and Mollaei-Renani, 2014), and promote the electrode-electrolyte interaction (Ma et al., 2022; Tan et al., 2018; Zhang et al., 2013). Also, atomic S (i.e., -C-S-C- bonds) lowers the interaction energies of carbon nanomaterials and facilitates the absorption of water. On the other hand, sulfur atom contains large polarizable d-orbitals, therefore, the lone pairs can easily interact with the molecules in the surrounding electrolyte (Li et al., 2015; Wohlgemuth et al., 2012b). In our previous study, low ionic strength of drinking water led to a significantly lower adsorption of antibiotics at the graphene sponge electrodes compared to the phosphate buffer solution of higher electrical conductivity (Ormeno-Cano and Radjenovic, 2022). As explained above, this was due to  the increased thickness of the electric double layer in low ionic strength solutions, which limits the electrostatic and ionic interaction between the target contaminants and the electrode surface (Liu et al., 2019; F. Wang et al., 2015). Addition of S-functional groups into the RGO coating improves the adhesion of all target contaminants due to an enhanced wettability and thus facilitated adsorption and ion bridging of organic contaminants with the electrode surface (Liu et al., 2019).

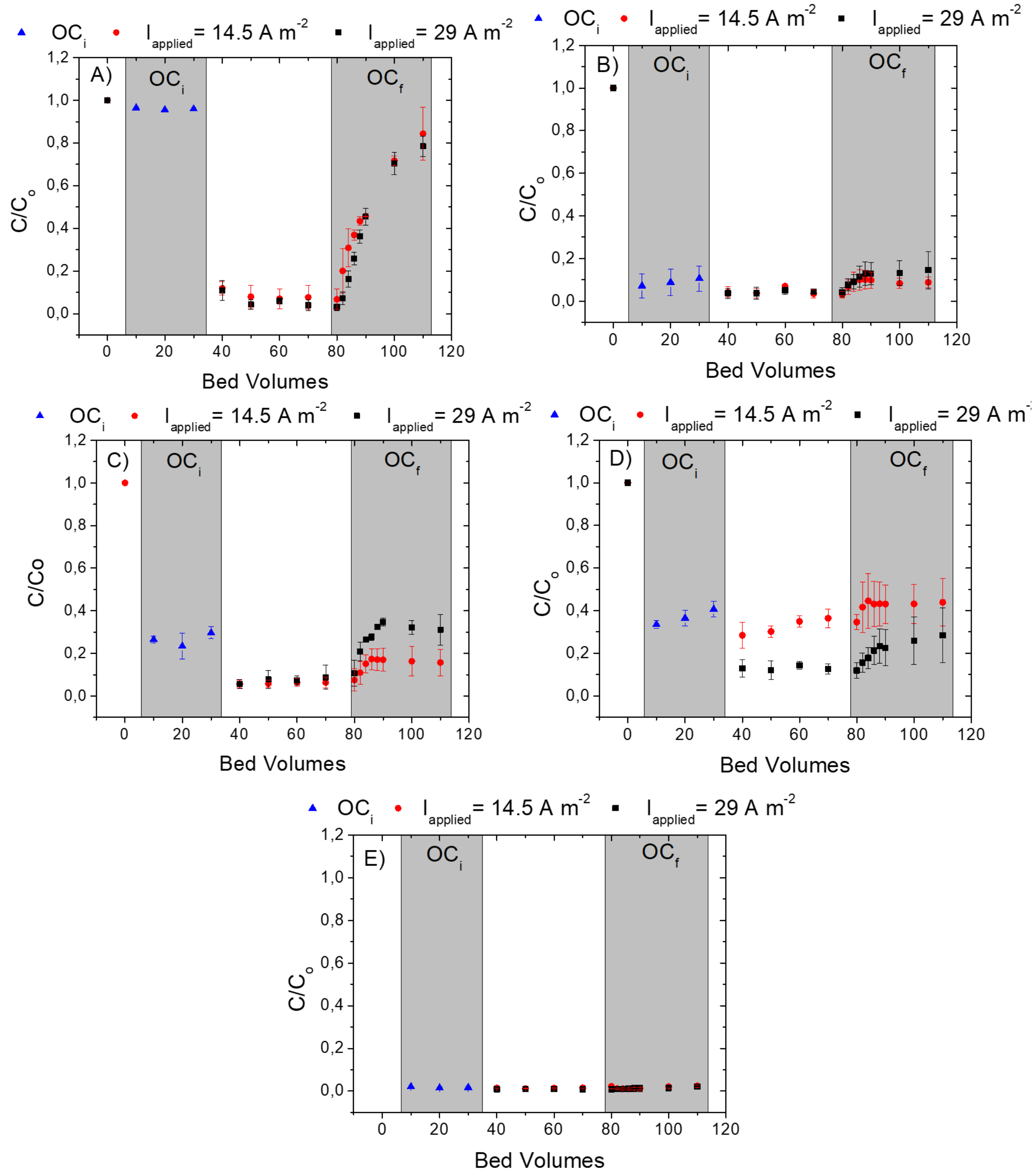

**Figure 4.** Measured effluent concentrations (C) of antibiotics normalized to their influent concentrations ($C_0$) of: **A)** SMX, **B)** ERT, **C)** ROX, **D)** TMP, and **E)** OFX for the SRGO/NRGO system at 14.5 and 29 A $m^{-2}$ of applied current density in drinking water.

Application of 14.5 and 29 A $m^{-2}$ of anodic current density increased the removal of SMX, ROX and TMP compared with the $OC_i$. In the case of poorly adsorbed SMX, complete removal was achieved already at the anodic current density of 14.5 A $m^{-2}$ (**Figure 4, Table S9**). This is significantly higher compared with the system using an undoped RGO anode (**Figure S5**), where 14.5 and 29 A $m^{-2}$ achieved 47% and 71% removal of SMX, respectively. Similar results were observed for TMP, with 42% and 49% removal at 14.5 and 29 A $m^{-2}$ in the RGO/NRGO system, whereas the SRGO/NRGO reactor achieved 67% and 87% removal, respectively. Both SMX and TMP have limited reactivities with ozone ($k_{O3}$ (SMX)=5.5x$10^5$ $M^{-1}s^{-1}$, and $k_{O3}$(TMP)=2.7x$10^5$ $M^{-1}s^{-1}$) (**Table S11**), and are easily oxidized by the electrogenerated •OH, with the bimolecular rate constants of 5.5x$10^9$ and 6.9x$10^9$ $M^{-1}s^{-1}$, respectively (**Table S11**). Given that the determined •OH concentrations were similar for the two employed anodes, improved removal of SMX and TMP when using the SRGO anode is likely a consequence of their improved interaction with the S-functionalized graphene sponge electrode, as observed also in the $OC_i$. S-functionalization also positively impacted the removal of ROX, with 37% (14.5 A $m^{-2}$) and 57% removal (29 A $m^{-2}$) in the RGO/NRGO system, and 92-94% removal in SRGO/NRGO system (**Table S9**). ERT and OFX were completely adsorbed already in the $OC_i$, when using the SRGO electrode, thus further enhancement in their removal when applying the current could not be observed. It should be noted that adsorption of trace organic contaminants onto the graphene sponge electrodes represents an advantage for practical applications, as they act as capture step and make the electrochemical degradation of contaminants less limited by the mass transfer. Given the pronounced adsorption of the target antibiotics in the SRGO/NRGO system, and near complete removal already at the lower current density applied (14.5 A $m^{-2}$), further increase in current density to 29 A $m^{-2}$ lead only to a somewhat higher removal efficiency in the case of TMP (i.e., from 67% to 87% removal, respectively). Furthermore, slower return of the effluent concentrations to the influent concentration values in the $OC_f$ in the SRGO/NRGO system (**Figure 4**) compared with the RGO/NRGO system (**Figure S5**) confirmed on one hand that there was no accumulation of the contaminants in the system, and on the other hand, it indicated an improved retention of charge of the SRGO electrode. Higher capacitance of the SRGO electrode enhances the electro-sorption and electro-degradation of the contaminants for a longer period after the current is switched off compared with the RGO electrode. The energy consumption of the SRGO/NRGO system treating low-conductivity drinking water at 14.5 A $m^{-2}$ was 0.43 kWh $m^{-3}$ and yielded ≥94% removal of all target antibiotics apart from TMP (67%). $E_{EO}$ was in the range 0.76±0.4 kWh $m^{-3}$ (OFX) to 2.8±0.74 kWh $m^{-3}$ (ROX) (**Table 1, Figure S6**).

**Table 1.** Electric energy per order of magnitude ($E_{EO}$) (kWh $m^{-3}$) of SRGO/NRGO system at 14.5 and 29 A $m^{-2}$ of applied current density in drinking water.

| Antibiotic | (%) Removal **I= 14.5 A $m^{-2}$** | $E_{eo}$ (kWh $m^{-3}$) | (%) Removal **I= 29A $m^{-2}$** | $E_{eo}$ (kWh $m^{-3}$) |
|---|---|---|---|---|
| SMX | 94.4 ± 3.11 | 1.31±0.25 | 91.7±2.11 | 2.5± 0.19 |
| ERT | 95.7 ± 1.56 | 1.32±0.33 | 95.9±0.51 | 1.7 ±0.12 |
| ROX | 93.6 ± 0.72 | 0.95±0.09 | 92±1.8 | 2.8±0.74 |
| TMP | 67.1± 3.41 | ** | 87.2±0.93 | ** |
| OFX | 98.5 ±0.39 | 0.76±0.08 | 99.2±0.1 | 0.8±0.02 |

** one order of removal was not achieved.

## Conclusions

The introduction of sulfur-based functional groups such as -C-SOx-C- and -C-S-C- groups into the RGO coating enhanced the hydrophilicity and electrochemically active surface area of the graphene sponge electrode, and thus improved the electrode wetting and interaction of the electrode surface with the target organic and microbial contaminants. S-functionalized graphene sponge electrode enhanced the inactivation of the multi-drug resistant *E. coli* in drinking water from 1.7 log (RGO) to 4.5 log (SRGO) at 29 A $m^{-2}$ of applied current density, and with an energy consumption of 1.1 kWh $m^{-3}$. Storage experiments confirmed no regrowth of *E. coli* and yielded their further inactivation, evidencing an irreparable damage of the bacterial cell walls caused by the electrochemical treatment, and resulting in the overall *E. coli* removal of 5.2 log. S-functionalization also significantly improved the adsorption of antibiotics onto the electrode surface in low conductivity drinking water, which enabled their complete removal and further electrochemical degradation at low current densities, resulting in low energy consumption (1.1 kWh $m^{-3}$).

Currently, high purity graphene oxide is sold at prices of €1,200-1,500/kg GO. The employed loading of GO was ~0.05 g of GO per g of mineral wool (obtained loading: 0.038 g of RGO per g of mineral wool), thus indicating an estimated material cost of €0.08 per gram of mineral wool, calculated for the GO loading employed in excess. The cost of the mineral wool was not considered as it is negligible (less than 50 ¢/kg mineral wool). Thus, this would result in a cost of €46 per $m^2$ of the projected electrode surface area, which makes graphene-based sponges developed in this study very cost-competitive compared with the state-of-the-art BDD anodes with an approximate cost of €6,000 per $m^2$. Given that the developed anode material does not rely on the production of chlorine for electrochemical disinfection, the formation of disinfection byproducts is effectively avoided. Inherent characteristics of electrochemical systems such as operation at ambient temperature and pressure, modular design, no usage of chemicals and easy coupling with renewable energy, as well as the low-cost of the developed graphene sponge electrodes (<50 € per $m^2$), make this an attractive technology for point-of-use removal of antibiotics and ARBs disinfection of drinking water.

## Acknowledgments

The authors would like to acknowledge ERC Starting Grant project ELECTRON4WATER (grant number 714177). ICRA researchers thank funding from CERCA program. Moreover, the authors acknowledge the support from the Economy and Knowledge Department of the Catalan Government through a Consolidated Research Group (ICRA-TECH - 2021 SGR 01283). N.O. thanks José Luis Balcazar and Carles Borrego for providing the *E. coli* strain (DSM 103246).

# Supplementary Material

# Electrochemical removal of antibiotics and multi-drug resistant bacteria using S-functionalized graphene sponge electrodes

*Natalia Ormeno-Cano*[a,b], *Jelena Radjenovic* [a,c*]

[a]*Catalan Institute for Water Research (ICRA- CERCA), c/Emili Grahit, 101, 17003 Girona, Spain*

[b]*University of Girona, Girona, Spain*

[c]*Catalan Institution for Research and Advanced Studies (ICREA), Passeig Lluís Companys 23, 08010 Barcelona, Spain*

* *Corresponding author:*

*Jelena Radjenovic, Catalan Institute for Water Research (ICRA), Scientific and Technological Park of the University of Girona, 17003 Girona, Spain*

Phone: + 34 972 18 33 80; Fax: +34 972 18 32 48; E-mail: jradjenovic@icra.cat

**Text S1**. **Characterization of synthesized graphene-based sponges.**

Scanning electron microscopy (SEM) was carried out with a FEI Quanta FEG (pressure: 70Pa; HV: 20kV; and spot: 4). The X-ray photoelectron spectroscopy (XPS) measurements were performed with a Phoibos 150 analyzer (SPECS GmbH, Berlin, Germany) in ultra-high vacuum conditions (base pressure 1-10 mbar) with a monochromatic aluminium Kalpha x-ray source (1486.74 eV). The energy resolution as measured by the FWHM of the Ag 3d5/2 peak for a sputtered silver foil was 0.58 eV. X-ray diffraction (XRD) analysis was conducted with an X'pert multipurpose diffractometer using a Cu Kα radiation (l = 1.540 Å), at room temperature. The apparatus was equipped with a vertical θ–θ goniometer (240 mm radius) with fixed sample stages that do not rotate around the Ω axis as in the case of Ω–2θ diffractometers. An ultrafast X-ray detector based on real-time multiple strip technology was utilized as an X'Celerator. The diffraction pattern was recorded with a step size of 0.03°C and a time per step of 1,000 seconds, between 4°C and 30ºC.  Zeta (ζ) potential of SRGO was measured using Zetasizer Nano ZS (Malvern Panalytical Ltd) operating with a 633 nm laser and using an aqueous solution at pH 7. The contact angle of the synthesized graphene sponges was determined with a DSA25S Drop shape analyzer (KRÜSS Scientific, Hamburg, Germany), using the sessile drop technique and the ADVANCE software for analysing the contact angle of the water drops.

**Text S2. Calculation of the electric energy per order ($E_{eo}$, kWh $m^{-3}$), energy consumption ($E_c$, kWh $m^{-3}$) and electrochemically active surface area (EASA).**

In the case of one order of magnitude of removal, electric energy per order ($E_{eo}$, kWh $m^{-3}$) was calculated according to the formula:

$$E_{eo} = \frac{U.I}{\mathrm{q}\,.\,\log\frac{C_0}{C}} \qquad \text{(eq. 1)}$$

where $q$ is the flow rate (L $h^{-1}$), $U$ the average cell voltage (V) and $I$ the applied current (A) and $C_0$ and C are the influent and effluent contaminant concentrations (µM, note that for an order of magnitude removal, log $C/C_0$ =1). For a removal of less than one order of magnitude, the energy consumption was estimated using (eq 2) as follow:

$$E_c = \frac{U * I}{\mathrm{q}} \qquad \text{(eq.2)}$$

Electrochemically active surface area (EASA) was calculated according to the equation:

$$\mathrm{EASA} = C_{dl}/C_d \qquad \text{(eq. 3)}$$

where $C_{dl}$ represent the imaginary part of the double-layer capacitance ($C_{dl}$) and $C_d$ is a constant value of carbon-based electrode (10 µF $cm^{-2}$) (Cuervo Lumbaque et al., 2022).

**Text S3. Calculation of the ohmic drop.**

The internal resistance was measured through the electrochemical impedance spectroscopy (EIS) experiments of SRGO(A)-NRGO(C) and RGO(A)-NRGO(C) systems and were used to calculate the ohimc drop. An equivalent circuit as the one presented in image (1) was employed to predict the internal resistances of our system. The internal circuit consisting of the ohmic internal resistance (R1) connected in series with a charge transfer resistance (R2) and the capacitance (C2), both connected in parallel to each other. (Q3) represents a constant phase element indicating the double layer capacitance, which occurs at the interface between the material and the electrolytes due to charge separation, and is connected in parallel with (Q4), the ideal polarizable capacitance and (R3) a charge separation resistance.

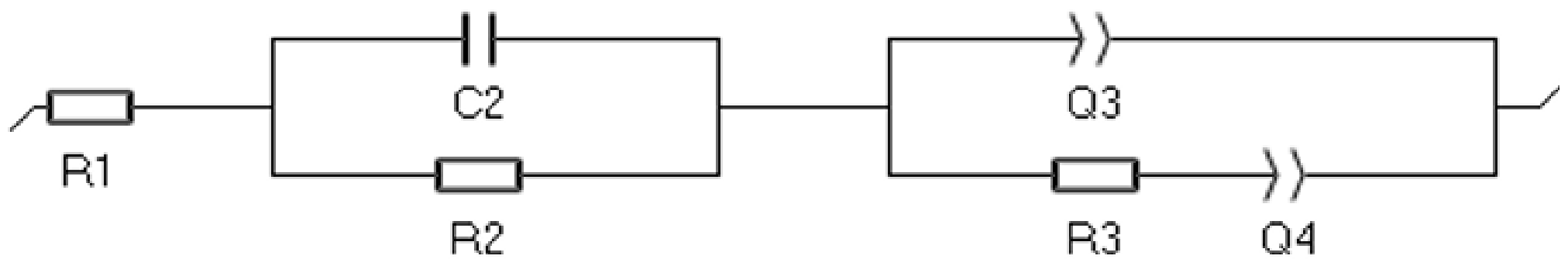

(Img.1)

The ohmic drop-corrected anode potential ($E_{corr}$, V) was calculated according to the following equation:

$$E_{corr} = E_{rec} - (I \times R1) \quad \text{(eq. S1)}$$

where $I$ is the applied anodic current (A), $R1$ is the uncompensated resistance (Ω) and $E_{rec}$ is the anode potential recorded in the chronopotentiometric experiments (V).[1]

**Text S4**. **Determination of the electrogenerated oxidant species ($H_2O_2$, $O_3$ and ·OH).**

The concentrations of ozone and $H_2O_2$ in the effluent samples were determined immediately after sampling. Ozone ($O_3$) was measured (in the absence of chlorine) using N,N-diethyl-p-phenylenediamine (DPD) colorimetric method (Chlorine/Ozone/Chlorine dioxide cuvette tests LCK 310, Hach Lange, Spain). $H_2O_2$ was measured with a spectrophotometric method, using 0.01 M $CuSO_4$ solution and 0.1% w/v 2,9 dimethyl-1,10-phenanthroline (DMP) solution, based on the formation of $Cu(DMP)^{2+}$ that shows an absorption maximum at 454 nm (Baga et al., 1988). The steady-state concentration of ·OH was evaluated using terephthalic acid (TA) as a probe compound, due to tis high reactivity with ·OH, resistance to direct electrolysis (Guo et al., 2016) and poor reactivity with $O_3$ (Zang et al., 2009). The determined concentration of ·OH should be considered as a lower bound of the steady-state concentration of electrochemically generated ·OH (Guo et al., 2016). TA was measured using high performance liquid chromatography coupled to ultraviolet detector (HPLC-UV) (Agilent Technologies 1200 series) at 239 nm (quantification limit of 1.98 mg $L^{-1}$).

**Text S5. Analysis of target antibiotics.**

Target antibiotics were analyzed using an ultraperformance liquid chromatography (UPLC, Waters, USA) with an Acquity UPLC HSS T3 column (2.1×50 mm, 1.8 µm, Waters), coupled to a quadrupole linear ion trap mass spectrometer (QqLIT-MS) with a turbo Ion Spray source (5500 QTRAP, Applied Biosystems, USA).Roxythromycin (ROX), erythromycin (ERT), trimethoprim (TMP) Sulfamethoxazole (SMX), and ofloxacin (OFX), were analysed in electrospray (ESI) positive mode using an Acquity ultraperformance liquid chromatography (UPLC) HSS T3 column (2.1×50 mm, 1.8 µm, Waters) run at 30°C. As it was described in our previous study, acetonitrile with 0.1% formic acid (eluent A), and milli-Q (LC-MS grade) water with 0.1% formic acid (eluent B) at a flow rate of 0.5 mL $min^{-1}$were employed organic and liquid phases. The gradient was started at 2% of eluent A that was increased to 70% A by 4 min, further increased to 95% A by 5 min and it was kept constant by 5.5 min. Then, it was returned to the initial conditions of 2% A by 6 min further kept constant for 1 min. The total run time was 7 min. The target antibiotics were analysed in a multiple reaction monitoring (MRM). The settings for the compound-dependent parameters of each transition are summarized in **Table S2**. The source-dependent parameters were as follows: curtain gas (CUR), 30 V; nitrogen collision gas (CAD), medium; source temperature (TEM), 650°C; ion source gases GS1, 60 V and GS2, 50 V; ion spray voltage, 5500V, and entrance potential (EP), 10V.[1]

**Table S1.** Drinking water characteristics.

| | |
|---|---|
| **Conductivity (mS cm$^{-1}$)** | 0.450 |
| **Total alkalinity ($T_{AC}$) mg L$^{-1}$ $CaCO_3$** | 130.1 |
| **$Cl^-$, mg L$^{-1}$** | 21.4 |
| **$SO_4^{2-}$, mg L$^{-1}$** | 13.4 |
| **$Na^+$, mg L$^{-1}$** | 13.8 |
| **$Mg^{2+}$, mg L$^{-1}$** | 9.2 |
| **$Ca^{2+}$, mg L$^{-1}$** | 51.2 |
| **Total organic carbon (TOC), mg L$^{-1}$** | 2.1 |
| **pH** | 7.3 |

**Table S2.** The optimized compound-dependent MS parameters: declustering potential (DP), collision energy (CE), and cell exit potential (CXP), for each compound and each transition of the negative and positive mode.

| Organic compound | Q1 Mass (Da) | Q3 Mass (Da) | DP | CE | CXP |
|---|---|---|---|---|---|
| TMP | 291.02 | 230.2 | 91 | 33 | 12 |
| | 291.021 | 261.1 | 91 | 35 | 10 |
| SMX | 253.992 | 156.1 | 81 | 23 | 12 |
| | 253.992 | 92.0 | 81 | 37 | 12 |
| ERT | 734.3 | 576.4 | 116 | 27 | 22 |
| | 734.3 | 158.1 | 116 | 39 | 14 |
| OFX | 362.038 | 318.3 | 86 | 27 | 12 |
| | 362.038 | 261.0 | 86 | 39 | 12 |
| ROX | 837.37 | 158.1 | 91 | 31 | 26 |
| | 837.37 | 158.1 | 91 | 43 | 14 |

**Table S3.** XPS atomic content of synthesized RGO and SRGO sponges.

| | SRGO | RGO [2] |
|---|---|---|
| **C (%)** | 68.2 | 70.3 |
| **O (%)** | 30.6 | 28 |
| **S (%)** | 1.3 | 0 |
| **C/O** | 2.23 | 2.51 |

**Table S4.** XPS atomic content of the SRGO sponge with the % of functional groups of C1s, O1s and S2p XPS spectra.

| **C1s (%)** | | | | **O1s (%)** | | | **S2p (%)** | | | |
|---|---|---|---|---|---|---|---|---|---|---|
| ***C=C 284.4eV*** | ***C-O/C-S 286.1eV*** | ***C=O 287.8eV*** | $\pi$-$\pi$ ***290.5eV*** | ***C-O 531.3eV*** | ***C=O 533.3eV*** | ***O=C-O 534.3eV*** | ***-C-S-C- 163.3eV*** | ***-C-SOx-C- (x=2,3,4) 168.39eV*** | ***-C-SOx-C- (x=2,3,4) 170.5eV*** | ***-C-SOx-C- (x=2,3,4) 172.8eV*** |
| 33.3 | 41.9 | 20.3 | 4.4 | 23.9 | 14.8 | 61.3 | 7.33 | 41.9 | 42.4 | 8.4 |

**Table S5.** Recorded total cell potentials ($E_{tot}$, V) and ohmic-drop corrected anodic potential ($E_{corrected}$, V) for RGO and SRGO at 14.5 and 29 A $m^{-2}$ of applied current density in drinking water.

| RGO | 14.5 A $m^{-2}$ | 29 A $m^{-2}$ |
|---|---|---|
| $E_{anode}$ (V) | 1.9 | 2.4 |
| $E_{cathode}$ (V) | -3.7 | -5.1 |
| $E_{total}$ (V) | 5.6 | 7.5 |
| $\Delta E_t$ (V/cm) | 2.8 | 3.8 |
| $R_\Omega$ (Ω) | 13.5 | |
| $R_{ct}$ (Ω) | 48.7 | |
| $\mathbf{E_{corrected}}$ (V) | 1.6 | 1.7 |

| SRGO | 14.5 A $m^{-2}$ | 29 A $m^{-2}$ |
|---|---|---|
| $E_{anode}$ (V) | 3.3 | 3.8 |
| $E_{cathode}$ (V) | -1.9 | 3.3 |
| $E_{total}$ | 5.2 | 7.1 |
| $\Delta E_t$ | 2.6 | 3.6 |
| $R_\Omega$ | 13.7 | |
| $R_{ct}$ | 79 | |
| $\mathbf{E_{corrected}}$ | 2.9 | 3.1 |

**Table S6.** Removal of *E. coli* from drinking water using RGO and SRGO anodes coupled to NRGO cathode, at 14.5 and 29 $Am^2$ of applied current density.

| Log reduction | RGO | SRGO |
|---|---|---|
| **$OC_i$-** | 1.9 ± 0.18 | 1.9± 0.18 |
| **14.5 A $m^{-2}$** | 2.3 ± 0.13 | 3.7± 0.14 |
| **29A $m^{-2}$** | 1.7± 0.19 | 4.5± 0.16 |

**Table S7.** Measured concentrations of oxidant species determined in anode-cathode ($^{\bullet}OH$, $H_2O_2$) and cathode-anode ($O_3$) flow direction at 14.5 and 29 A $m^{-2}$ of applied current density in drinking water.

| | RGO | | SRGO | |
|---|---|---|---|---|
| **I (A $m^{-2}$ )** | **14.5** | **29** | **14.5** | **29** |
| $O_3$ (mg $L^{-1}$) | 0.1±0.04 | 0.16±0.013 | 0.07 ±0.01 | 0.5±0.01 |
| $H_2O_2$ (mg $L^{-1}$) | 0.64 ±0.01 | 0.65±0.03 | 0.54±0.02 | 0.58±0.04 |
| OH• (M) | $4.10^{-14} \pm 1.3\ 10^{-15}$ | $4.69\ 10^{-14} \pm 1.9\ 10^{-15}$ | $3.1\ 10^{-14} \pm 9.4\ 10^{-16}$ | $4.10^{-14} \pm 5.10^{-15}$ |

**Table S8.** Comparison of studies focusing on the electrochemical inactivation of different pathogens.

| Anode material | Operational mode | Electrolyte | Pathogen | Current density (mA $cm^{-2}$) | Energy consumption (KWh $m^{-3}$) | Log removal | Study |
|---|---|---|---|---|---|---|---|
| **SRGO, S-doped reduced graphene oxide** | Flow-through continuous mode | Drinking water | *E. Coli* DSM103246 | 50 | 1.1 | 5.2 log | This study |
| **Magnéli phase Ti4O7** | Flow-through continuous mode | 0.05 M Na2SO4 | *E. Coli* | 5 | 1.26 | 4.9 log | [1] |
| **TiO2 REM** | Batch mode | 10 mM NaClO4 | *E. Coli* | 0.03 - 0.50 | 2.0-88 | No live bacteria | [2] |
| **Magneli phase Ti4O7** | Batch mode | 0.05 M $Na_2SO_4$ | *Salmonella enterica serotype Typhimurium DT104* | 10.0 | 2.3 | 6.2 log | [3] |
| **$Mo_2C$** | Batch mode | PBS | *E. coli* K-12 LE392 | 4.04 | 0.27 | 5 log | [4] |
| **Pt clad Nb mesh** | Flow-through Continuous mode | 0.03 M $Na_2SO4$ | *E. coli* ATTC15597 | 32 | 6.3 | 4 log | [5] |
| **Mixed Metal Oxide (MMO)** | Flow-through continuous mode | synthetic hospital urine | *K. pneumoniae* | 50 | 1.1 | 7 log | [6] |
| **3D foam oxide nanowire (CuON)** | Flow-through continuous mode | 9 $gL^{-1}$ NaCl | *E. coli*ATTC15597), *Enterococcus faecalis* (ATCC 19433), and *Bacillus subtilis* (ATCC 6633) | 5.4 | $7.10^{-3}$ | >7 log | [7] |

**Table S9.** Removal efficiency (%) of antibiotics in the initial open circuit conditions ($OC_i$) and at 14.5 and 29 A $m^{-2}$ of anodic current density in drinking water.

| ANTIBIOTICS | SRGO/NRGO | | | RGO/NRGO | | |
|---|---|---|---|---|---|---|
| | **$OC_i$** | **14.5 A $m^{-2}$** | **29 A $m^{-2}$** | **$OC_i$** | **14.5 A $m^{-2}$** | **29 A $m^{-2}$** |
| ROX | 73.3±2.0 | 93.6±0.7 | 92±1.8 | 25±1.0 | 36.6.10±1.5 | 57.1±1.05 |
| TMP | 63.1±3.6 | 67.1±3.4 | 87.2±0.9 | 35.4±6.7 | 42.3±4.6 | 48.6±4 |
| ERT | 91.6±1.7 | 95.7±1.5 | 95.9±0.5 | 35.7±0.5 | 50.2±3 | 60.7±1.6 |
| OFX | 98.3±0.24 | 98.5±0.4 | 99.2±0.1 | 84.5±2.1 | 88.9±1.3 | 87±0.1 |
| SMX | 4±0.4 | 91.7±2.1 | 94.4±3.1 | 4.2±1.7 | 46.5±4.2 | 70.7±3.1 |

**Table S10.** Chemical structures and physico-chemical properties of the target contaminants; molecular weight (MW), pKa, octanol-water distribution coefficient calculated based on chemical structure at pH 7.4 (ACD/logD), and polar surface area. Calculated ACD/logD values and polar surface areas were collected from Chemspider.com database.

| **Organic compound (MW, g/mol)** | **Chemical structure** | **pKa** | **ACD/LogD** | **Polar surface area ($Å^2$)[1]** |
|---|---|---|---|---|
| Erythromycin (ERT)<br>733,93 g/mol | 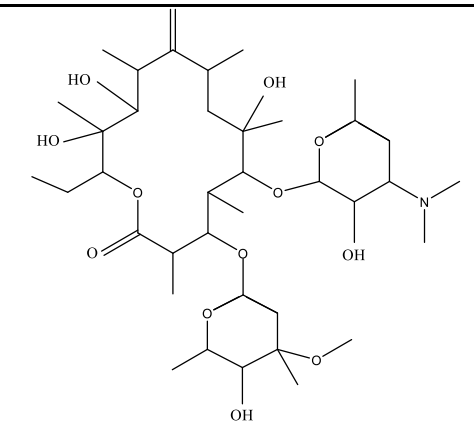 | 8.92 [8] | 1.69 | 194 |
| Ofloxacin (OFX)<br>361,37 g/mol<br>$C_{18}H_{20}FN_3O_4$ | 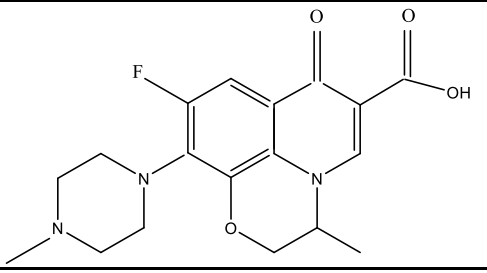 | $pK_{a1}$ = 5.97<br>$pK_{a2}$ = 9.28 [9] | -2.08 | 73 |
| Trimethoprim (TMP)<br>290,1 g/mol<br>$C_{14}H_{18}N_4O_3$ | 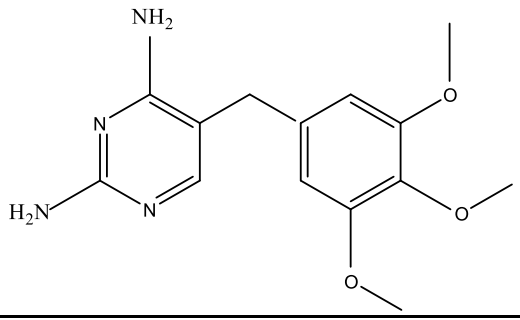 | $pK_{a1}$ =3.2[10]<br>$pK_{a2}$ =7.1 | -1.15 | 99 |
| Roxithromycin (ROX)<br>837.05 g/mol<br>$C_{41}H_{76}N_2O_{15}$ | 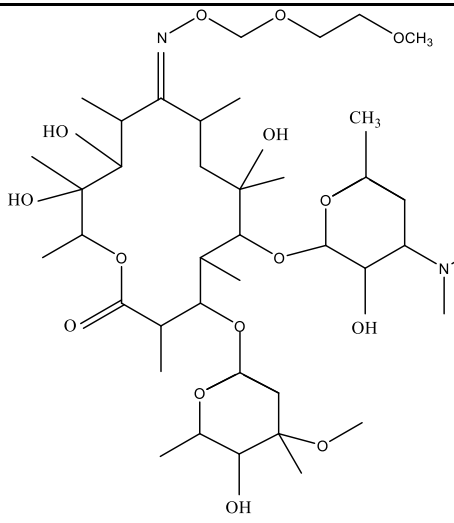 | 8.82[10] | 2.80 | 217 |
| Sulfamethoxazole (SMX)<br>253,28 g/mol<br>$C_{10}H_{11}N_3O_3S$ | 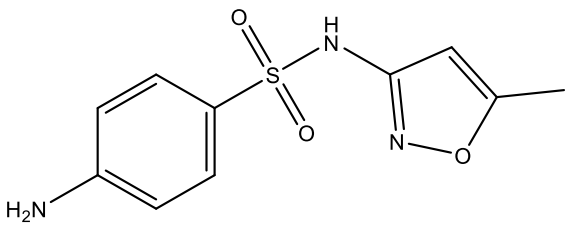 | $pK_{a1}$ =1.4[11]<br>$pK_{a1}$ =5.8 | -0.56 | 107 |

**Table S11**. Reported bimolecular rate constants of target antibiotics with ozone ($k_{O3}$, $M^{-1}$ $s^{-1}$) and hydroxyl radicals ($k_{OH}$, $M^{-1}$ $s^{-1}$).

| Compound | $k_{O3}$ ($M^{-1}$ $s^{-1}$) | $k_{OH}$ ($M^{-1}$ $s^{-1}$) |
|---|---|---|
| Erythromycin (ERT) $C_{37}H_{67}NO_{13}$ | $6.93x10^4$ [12] | $5x10^9$ [13] |
| Ofloxacin (OFX) $C_{18}H_{20}FN_3O_4$ | $2.0$ $x10^6$ [14] | $4.2$ $x10^9$ [14] |
| Trimethoprim (TMP) $C_{14}H_{18}N_4O_3$ | $2.7$ $x10^5$ | $6.9x10^9$ [15] |
| Sulfamethoxazole (SMX) $C_{10}H_{11}N_3O_3S$ | $5.5x10^5$ | $5.5x10^9$ [16] |
| Roxithromycin (ROX) $C_{41}H_{76}N_2O_{15}$ | $6.4x10^4$ [15] | $5.4x10^9$ |

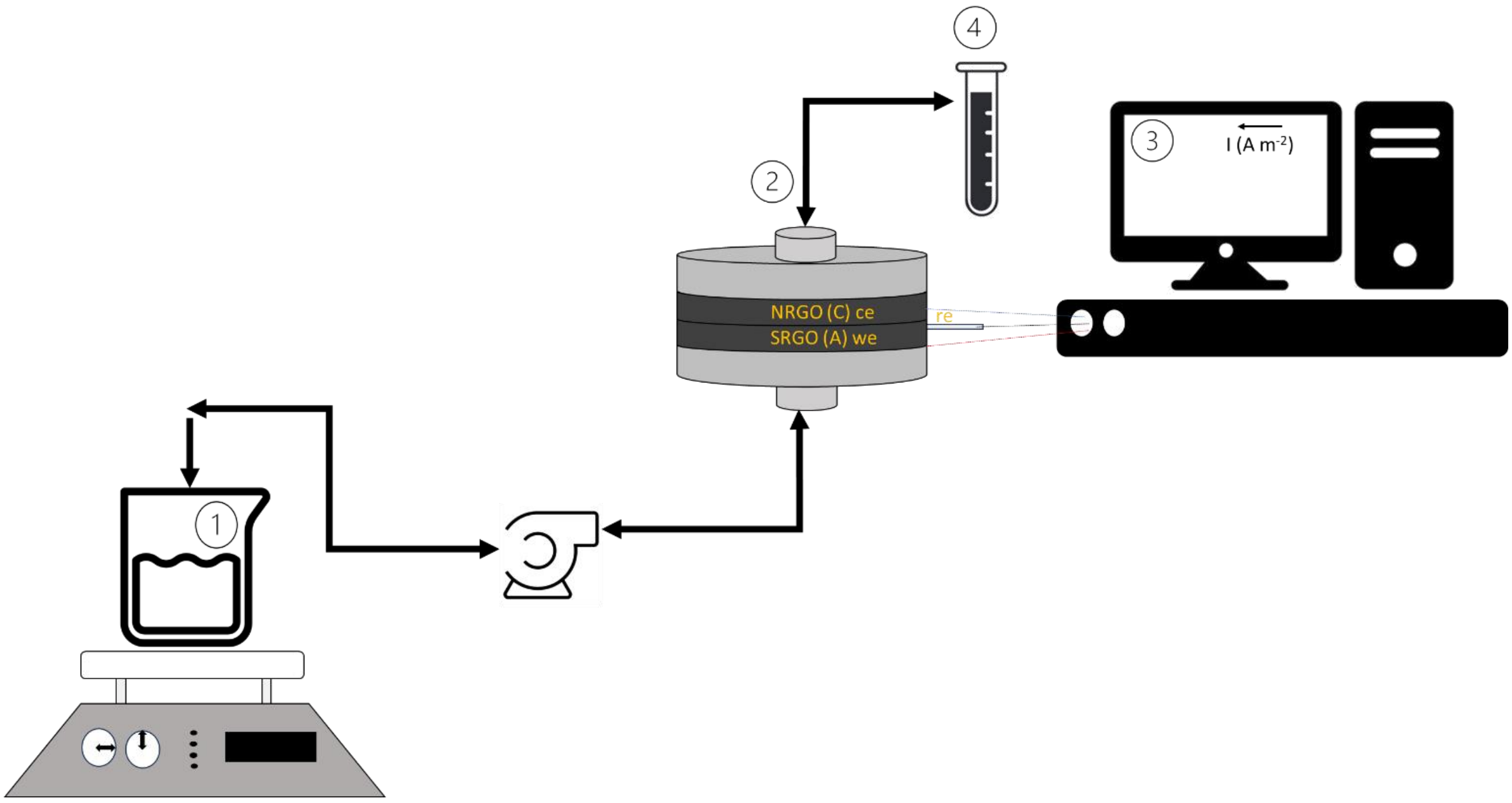

**Figure S1**. Scheme of the flow-through reactor and its experimental set-up employed (1) influent of drinking water spiked with a multi-drug resistant *E. coli* and antibiotics, (2) anode (working electrode, we), cathode (counter electrode, ce), reference electrode, (re, leak-free Ag/AgCl), (3) application of current to the anode through a potenciostat connected to a computer for acquiring all data. and (4) effluent.

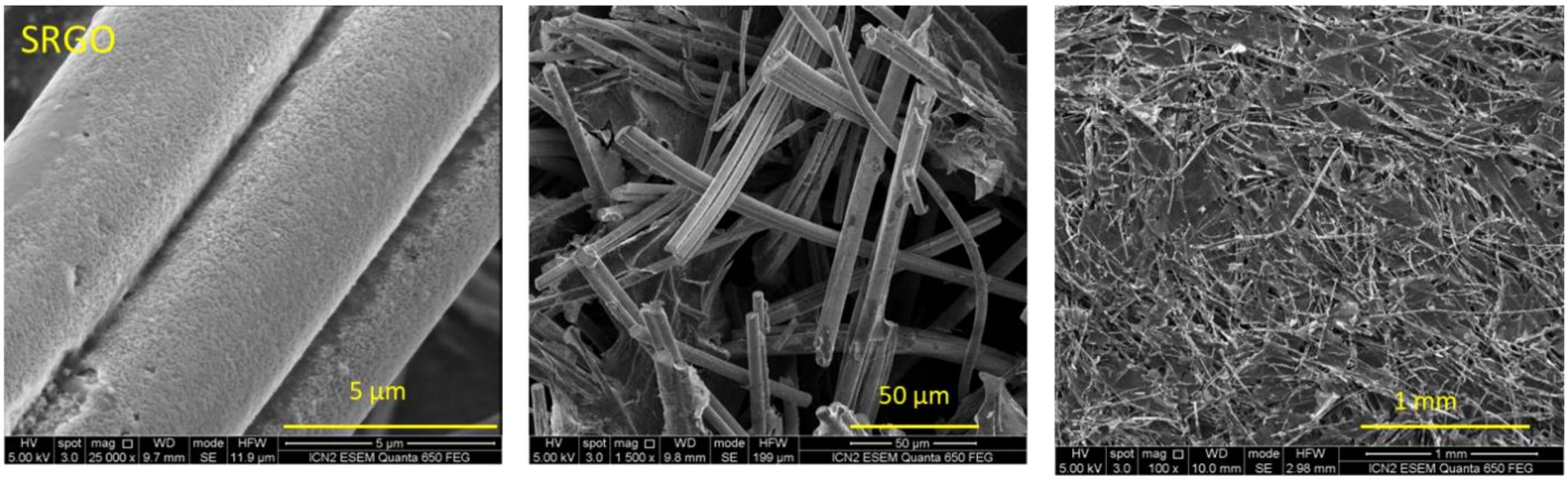

**Figure S2**. SEM images of the SRGO sponge**.**

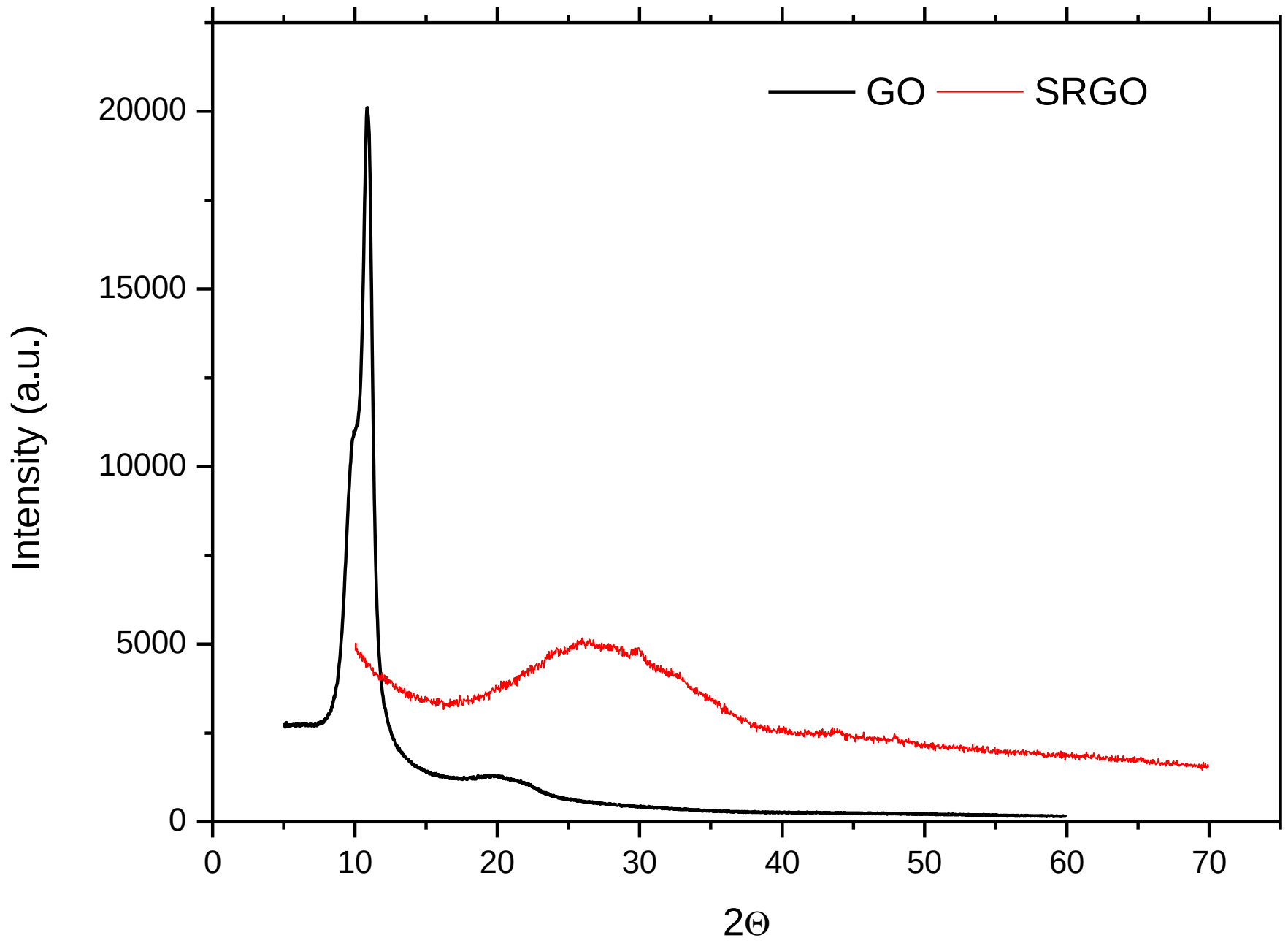

**Figure S3**. X-ray powder diffraction (XRD) patterns of GO and SRGO sponge.

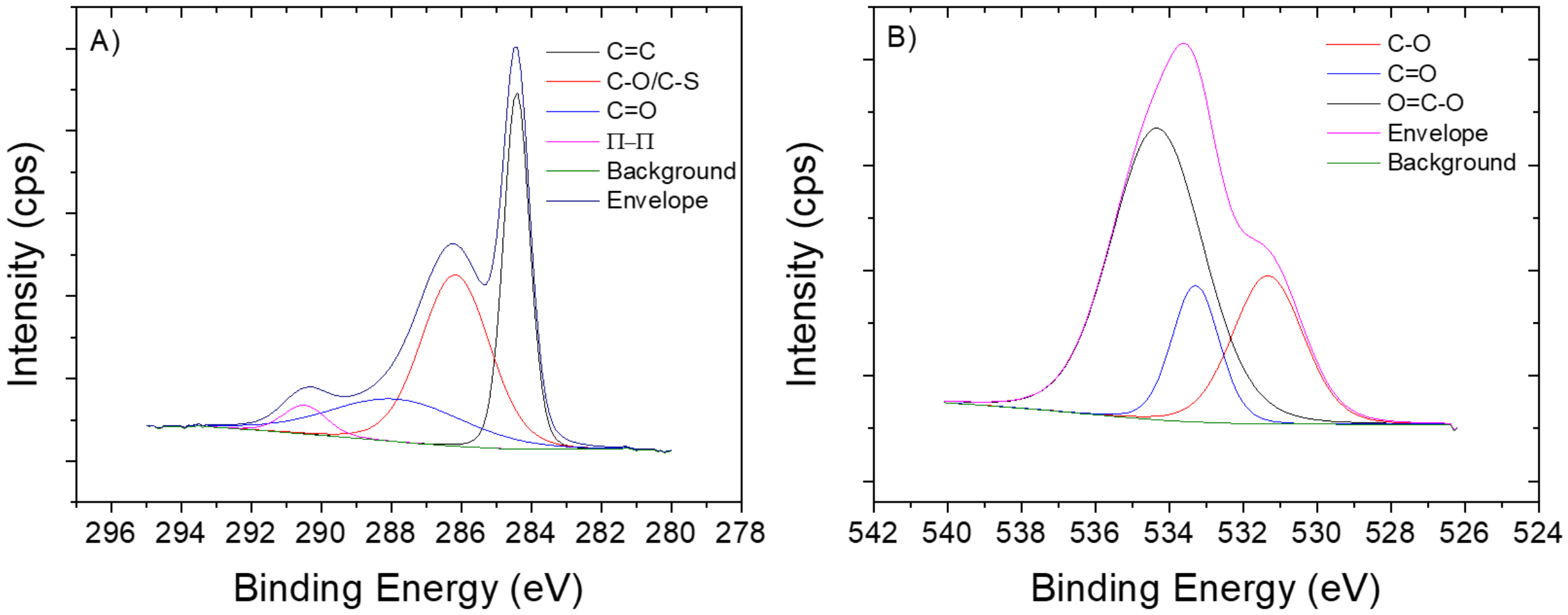

**Figure S4**. XPS spectra of: **A)** of C1s, and **B)** O1s of SRGO.

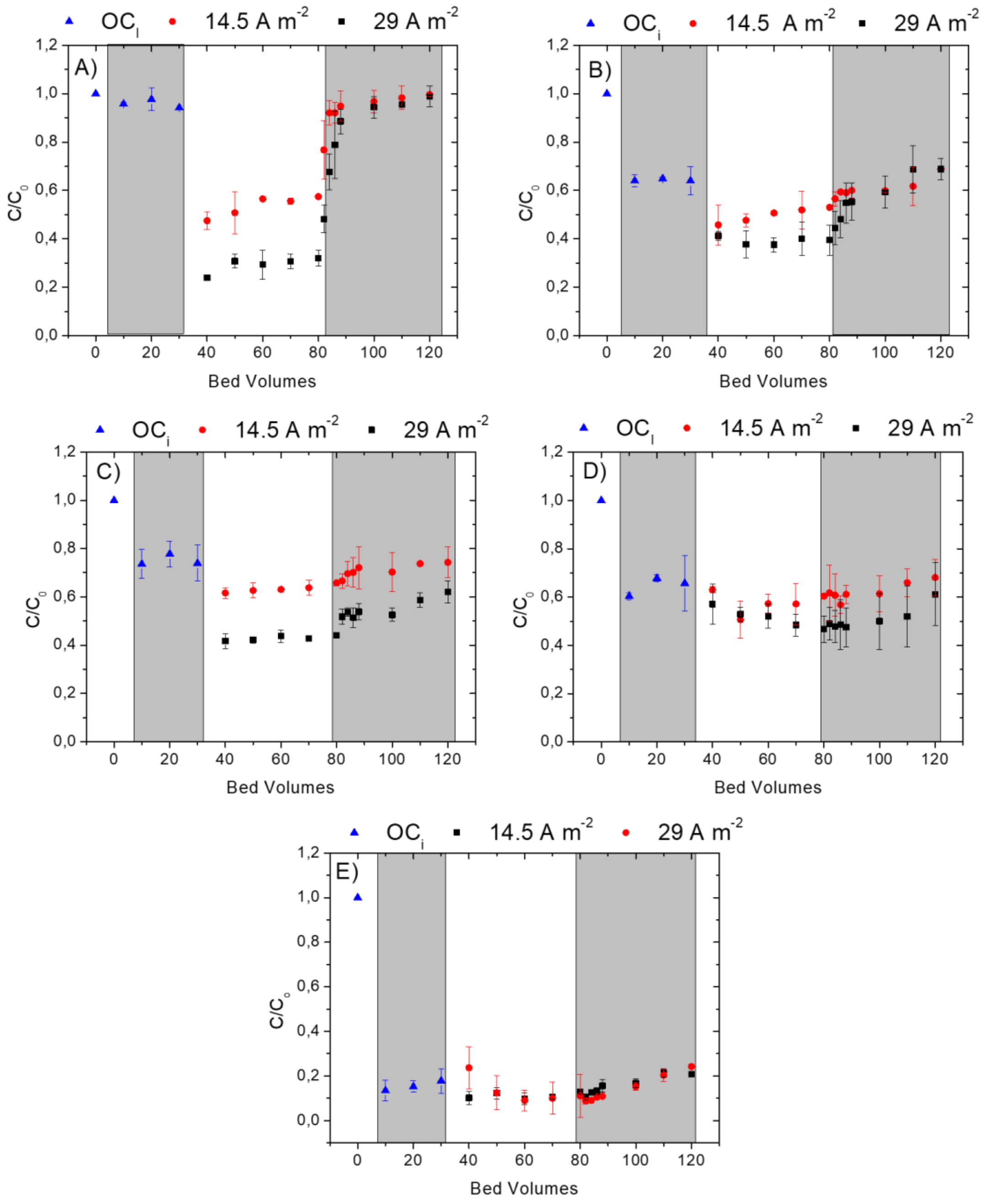

**Figure S5.** Measured effluent concentrations (C) of antibiotics normalized to their influent concentrations ($C_0$) of **A)** SMX, **B)** ERT, **C)** ROX, **D)** TMP, and **E)** OFX for the RGO/NRGO system at 14.5 and 29 A $m^{-2}$ of applied current density in drinking water.

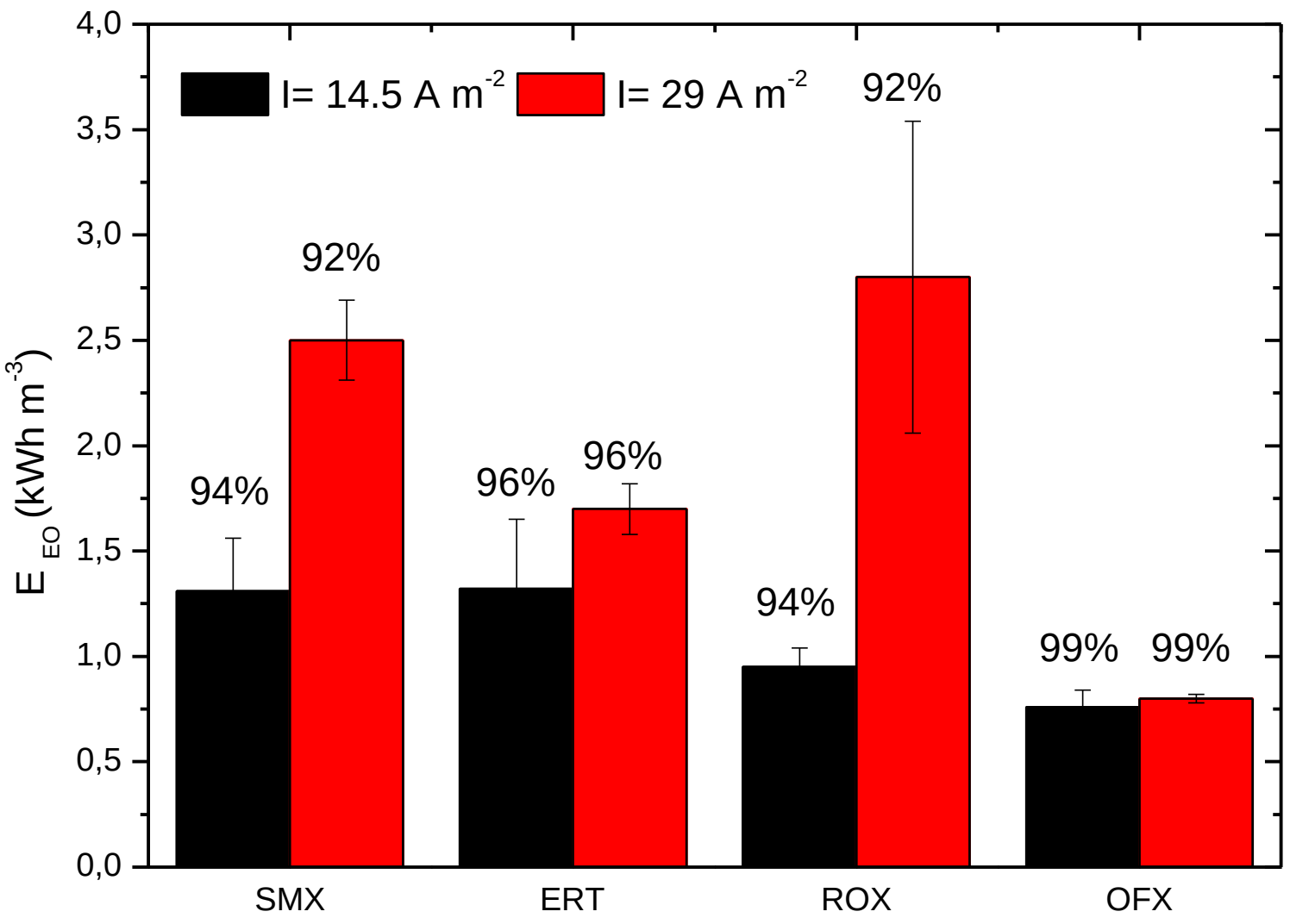

**Figure S6.** Electric energy per order of magnitude ($E_{EO}$) (kWh $m^{-3}$) of SRGO/NRGO system at 14.5 and 29 A $m^{-2}$ of applied current density in drinking water.
*TMP did not achieve one order of magnitude. The Energy consumption, ($E_c$) had resulted in 1.1 kWh $m^{-3}$ for the removal of 87.2%.